\documentstyle[epsfig,aps,preprint]{revtex}
\tightenlines  
\draft  
\newcommand{\Pbar}{\not{\!P}}

\newcommand{\be}{\begin{equation}}
\newcommand{\ee}{\end{equation}}
\newcommand{\ba}{\begin{eqnarray}}
\newcommand{\ea}{\end{eqnarray}}

\newcommand{\nsigma}{\mbox{\boldmath $\sigma$}}

\newcommand{\nkappa}{\mbox{\boldmath $\kappa$}}
\newcommand{\neta}{\mbox{\boldmath $\eta$}}

\newcommand{\nl}{{\bf      l}}
\newcommand{\nn}{{\bf      n}}

\newcommand{\np}{{\bf      p}}       
\newcommand{\nq}{{\bf      q}}
\newcommand{\nr}{{\bf      r}}         
\newcommand{\ns}{{\bf      s}}

\newcommand{\nJ}{{\bf      J}}

\newcommand{\nP}{{\bf      P}}

\begin{document}
\begin{titlepage}
\mbox{} 
\vspace*{2.5\fill} 
{\Large\bf 
\begin{center}
%
Analysis of polarized $^{16}$O$(\vec{e},e'\vec{p})$ observables within the
relativistic distorted wave impulse approximation
%
\end{center}
} 
\vspace{1\fill} 
\begin{center}
{\large 
M.C. Mart\'{\i}nez$^{1}$, J.R. Vignote$^{2}$, J.A. Caballero$^{1}$, T.W. Donnelly$^{3}$, 
E. Moya de Guerra$^{4}$, J.M. Ud\'{\i}as$^{2}$
}
\end{center}

\date{\today}

\begin{small}
\begin{center}
$^{1}${\sl 
Departamento de F\'\i sica At\'omica, Molecular y Nuclear \\ 
Universidad de Sevilla, Apdo. 1065, E-41080 Sevilla, SPAIN 
}\\[2mm]
$^{2}${\sl 
Departamento de F\'\i sica At\'omica, Molecular y Nuclear \\ 
Universidad Complutense de Madrid, E-28040  Madrid, SPAIN 
}\\[2mm]
$^{3}${\sl 
Center for Theoretical Physics, Laboratory for Nuclear Science 
and Department of Physics\\
Massachusetts Institute of Technology,
Cambridge, MA 02139, USA 
}\\[2mm]
$^{4}${\sl Instituto de Estructura de la Materia, CSIC, Serrano 123, E-28006 Madrid, SPAIN}

\end{center}
\end{small}

\kern 1. cm \hrule \kern 3mm

\begin{small}
\noindent
{\bf Abstract}
\vspace{3mm}

Recoil nucleon transferred polarization observables in coincidence quasielastic electron scattering are
studied within the relativistic distorted wave impulse approximation. Results for response 
functions and polarization asymmetries are discussed for proton knockout from $p_{1/2}$, $p_{3/2}$ and
$s_{1/2}$ shells in $^{16}$O. The impact of spinor distortion is examined by comparing
the fully relativistic calculation with results obtained by projecting out the negative-energy components.
In particular, a careful analysis of effects linked to the description of the 
bound and scattered relativistic nucleon wave functions is presented. 
The high sensitivity of some polarization observables to the
dynamical enhancement of the lower components, already shown within the relativistic plane wave
impulse approximation, is proven to be maintained in the relativistic
distorted wave approach. Semi-relativistic approaches based on the effective momentum approximation are
also studied. Finally, comparison with experimental data and a brief analysis of effects linked to 
medium modified form factors is presented.

\kern 2mm

\end{small}

\kern 2mm \hrule \kern 1cm
\noindent MIT/CTP\#
\end{titlepage}


\section{Introduction}


A very topical issue in nuclear physics at present
is the search for evidence of possible modification of the nucleon form factors inside the nuclear medium. 
A number of double polarized $(\vec{e},e'\vec{p})$ experiments have been proposed or carried out 
recently to measure polarization transfer asymmetries, motivated by the hope that such observables
may provide valuable information that can shed some light on this issue. Importantly, transferred polarization
observables have been identified as being ideally suited for such studies: they are believed to be the least
sensitive to most standard nuclear structure uncertainties and
accordingly to provide the best opportunities
for studying the nucleon form factors in the nuclear medium. Polarization transfer data have been reported
recently for the case of $^{16}$O$(\vec{e},e'\vec{p})^{15}$N in~\cite{Mal00} and for 
$^{4}$He$(\vec{e},e'\vec{p})^{3}$H in~\cite{Die01,Str02}. Although the
experimental uncertainties in both cases make it difficult to draw
unambiguous conclusions on the nucleon form factors inside the nuclei, the data 
in~\cite{Str02} do seem to favour such a possibility. Specifically,
this means that comparisons of measured polarization asymmetries with those computed using the best
currently available nuclear models for the states and operators involved in the coincidence reaction in fact
show disagreements, and that these can be removed by modifying the nucleon form factors in a reasonable way. 

Of course, what constitutes the ``best currently available nuclear models" must be judged carefully. In particular,
the kinematic regime where the measurements have been undertaken is at relatively high energy --- to make the
reaction sufficiently impulsive to be at all interpreted as a simple single-nucleon knockout reaction --- and
it is clear that relativistic effects in wave functions and operators are essential. So, for instance, 
the data in~\cite{Die01} disagree significantly with the standard non-relativistic calculations; however, this
cannot be taken as evidence for nucleon modifications, since one finds that the results are (not unexpectedly) 
much more in accord with a fully relativistic approach. Also recent data on induced polarization in 
$^{12}$C~\cite{Woo98} strongly support an analysis based on the fully relativistic formalism~\cite{Udi00}.
These results are not surprising since spin and relativity are intrinsically 
related, and hence one may {\it a priori} consider
the relativistic formalism to be better suited to describe polarization observables. 

Indeed, most electron scattering experiments performed in the last decade have involved 
energies and momenta high enough
to invalidate the non-relativistic approximations assumed within the standard non-relativistic
distorted wave impulse approximation (DWIA), {\em i.e.,} bound and scattered wave
functions given as solutions of the Schr\"odinger equation, and one-body current operator 
resulting from a non-relativistic reduction. In
the relativistic distorted wave impulse approximation (RDWIA), nucleon wave functions are
described by solutions of the Dirac equation with scalar and vector (S-V) potentials, and the
relativistic free nucleon current operator is used. 

Relativistic effects can be classified into two basic categories
according to their origin, namely kinematical and dynamical effects. The former 
are due to the truncation of the current operator within the non-relativistic approach, 
the latter, dynamical effects, come from the difference between the
relativistic and non-relativistic wave functions.
Here one may distinguish a dynamical depression of the upper component of the scattered
nucleon wave function in the nuclear interior (Darwin term) and a dynamical enhancement of the 
lower components, mainly that corresponding to the bound nucleon wave function.

So far, RDWIA calculations for cross sections and response functions at low and high missing 
momenta~\cite{Udi93,Udi95,Udi96,Udi99,Udi01} have clearly improved 
the comparison with experimental data over the previous non-relativistic approaches. 
Moreover, RDWIA also predicts larger spectroscopic factors which are more in accord with 
theoretical calculations which incorporate correlations~\cite{Udi93,Udi01}.  

Concerning the current operators, in some recent studies~\cite{Udi01lund,Ama96,Ama96bis,Ama98,Jesch} new so-called 
`semi-relativistic' approaches have been introduced to describe
$(e,e'p)$ reactions. Here the `semi-relativistic' current operators are obtained by expanding only in 
missing momentum over the nucleon mass while treating the transferred energy and momentum exactly. 
This new approach has been proven to retain important aspects of relativity, and hence its predictions,
compared with the standard DWIA, agree much better with the RDWIA calculations. 

Concerning dynamical effects,
the enhancement of the lower components of bound Dirac
spinors~\cite{Udi99,Udi01} (not present in the
semi-relativistic approaches) has been shown to play a crucial role in the description
of the interference $TL$ response and left-right asymmetry $A_{TL}$.
Meson exchange currents and the $\Delta$-isobar contribution have recently been 
analyzed in~\cite{Ama03,Far03} within the semi-relativistic approach, 
also showing very significant effects, particularly due to the $\Delta$, at 
large missing momentum $p\geq 300$ MeV/c.

In this paper we focus on the analysis of polarized 
A$(\vec{e},e'\vec{p})$B observables within the framework of the
RDWIA. Our aim is to study the role played by both kinematical and dynamical 
relativistic effects in a consistent description of the polarized responses and asymmetries. 
This work extends the previous analyses presented in~\cite{Cris1,Cris2}
within the plane wave approach, now including a realistic description of the final-state interactions
(FSI) through relativistic optical potentials. The magnitude of relativistic effects on various
transfer polarization observables is carefully examined,
disentangling the role played by the various ingredients that enter in the fully relativistic
formalism. In particular, we extend the study of~\cite{Cris1} where within RPWIA we demonstrated
the importance of the negative-energy components 
of the relativistic bound nucleon in the description of the polarized responses
and transferred polarization asymmetries. The RDWIA analysis performed here allows one
to examine also the dynamical enhancement of the lower components in the scattered Dirac wave functions and
moreover, makes it possible to carry out meaningful comparisons with measured observables.

Returning to the issue of potential medium modifications of the
nucleon form factors, the current study has the following goal: we
wish to explore a selected set of model ``variations on a theme'' of
the type discussed above. In all cases we choose only modeling that is,
within the context of the general relativistic approach being adopted,
consistent with what we know about initial- and final-state wave
functions and one-body electromagnetic operators. Since equally
acceptable relativistic potentials exist when obtaining the states and
since alternative descriptions of the current operators are likewise
acceptable, it is impossible at present to define what is ``the best''
model. Our goal is to explore these acceptable models and where the
resulting polarization observables differ with the choice of model to
ascribe these variations to a (minimal) theoretical
uncertainty. Needless to say, all of this is within the general
context of relativistic mean-field modeling and so the resulting
uncertainties are minimal in the sense that effects that go beyond the scope of
the modeling might increase the uncertainties. In the final analysis,
only if medium modification effects are larger than the uncertainties
we find here, and only if the uncertainties that arise from
ingredients not in the present model can ultimately be shown to be
small, will a convincing case be made for the necessity of having such
medium modification effects.

The paper is organized as follows: in Section 2 we briefly introduce the general formalism for
A$(\vec{e},e'\vec{p})$B reactions focusing on the relativistic distorted wave impulse approximation.
Within this context, we also introduce the projected approach, the effective momentum approximation
(EMA-noSV) and the use of semi-relativistic current operators. 
By comparing them one may get a clear image
of the importance of relativity in these processes. In Section 3 we present and discuss the
results, paying special attention to the polarized responses and transferred polarization asymmetries. 
Finally, in Section 4 we summarize our conclusions.


\section{Description of $A(\vec{e},e'\vec{p})B$ reactions}

\subsection{General formalism. RDWIA}

In this section we briefly review the general formalism needed to describe coincidence 
$(\vec{e},e'\vec{p})$ reactions. We consider plane waves for the incoming and outgoing electron 
(treated in the extreme relativistic limit) and the Born approximation (one virtual photon exchanged). 
When the incoming electron is polarized and the final nucleon polarization is measured,
the differential cross section can be written as~\cite{Bof96,Kel96,Pick87,Pick89,Ras89}
\begin{equation}
\frac{d\sigma}{d\varepsilon_ed\Omega_ed\Omega_F}=\frac{\sigma_0}{2}[1+\nP\cdot
\nsigma+h(A+\nP'\cdot\nsigma)]\, ,
\label{difcross1}
\end{equation}
where the variables $\{\varepsilon_e, \Omega_e\}$ refer to the scattered electron and 
$\Omega_F$ to the ejected nucleon. The term
$\sigma_0$ is the unpolarized cross section, $h$ is the incident
electron helicity, $A$ denotes the electron analyzing power, and
$\nP$ ($\nP'$) represents the induced (transferred) polarization. Note that both $\nP$ and $\nP'$
depend on the outgoing nucleon
polarization, but $\nP'$ only becomes accessible when the incoming electron beam is polarized.
The cross section in Eq.~(\ref{difcross1}) can also be written in terms of nuclear responses 
as follows:
\begin{eqnarray}
\frac{d\sigma}{d\varepsilon_ed\Omega_ed\Omega_F}
 &=& K\sigma_{M}f_{rec}^{-1} \left\{ v_L\left(R^{L}+R^{L}_{n}\widehat{S}_n\right )+
     v_T\left (R^{T}+R^{T}_n\widehat{S}_n \right) \right. \nonumber \\
&+& \left.
 v_{TL}\left[\left(R^{TL}+R^{TL}_{n}\widehat{S}_n\right)\cos\phi +
             \left(R^{TL}_{l}\widehat{S}_l+R^{TL}_{s}\widehat{S}_s\right)\sin\phi \right] \right.
\nonumber \\
&+& \left. 
v_{TT}\left[\left( R^{TT}+R^{TT}_{n}\widehat{S}_n\right)\cos2\phi +
            \left( R^{TT}_{l}\widehat{S}_l+R^{TT}_{s}\widehat{S}_s\right )\sin 2\phi \right] \right.  
\nonumber \\
&+& \left. 
h\left\{v_{TL'}\left[\left(R^{TL'}_{l}\widehat{S}_l+R^{TL'}_{s}\widehat{S}_s\right)\cos\phi
+\left(R^{TL'}+R^{TL'}_{n}\widehat{S}_n\right)\sin\phi\right] \right. \right.
\nonumber \\
& + & \left. \left.
v_{T}\left[R^{T'}_{l}\widehat{S}_l+R^{T'}_{s}\widehat{S}_s\right]\right\} \right\} \, , 
\label{responses}
\end{eqnarray}
where $\phi$ is the azimuthal angle that
determines the outgoing nucleon momentum. The term $K$ is a kinematical factor given by
$K=p_FM_NM_B/M_A$, with $p_F$ the outgoing nucleon momentum, $M_N$ the nucleon mass, and
$M_B$ ($M_A$) the mass of the residual nucleus (target), respectively. 
The Mott cross section is represented by $\sigma_M$, $f_{rec}$ is the recoil factor given by $f_{rec}=1+(\omega p_F-qE_F\cos\theta_F)/M_Ap_F$, where $E_F$ is the outgoing nucleon energy and $\theta_F$ is the angle between $\np_F$ and the transferred momentum, and the $v_K$,
$K=L$, $T$, ... are the standard electron scattering kinematical factors (see~\cite{Ras89,Don86}).
The indices {$l$, $s$, $n$} refer as usual to the directions selected to specify the recoil
nucleon polarization: $\nl$ (parallel to the momentum 
$\np_F$), $\nn$ (perpendicular to the plane 
containing $\np_F$ and the transfer momentum $\nq$), and $\ns$ 
(determined by $\nn\times \nl$). From this large number of possible response functions
some selection can be made to limit the focus:
\begin{itemize}
\item Assuming coplanar kinematics, {\em i.e.,} $\phi=0^{\circ},180^{\circ}$, from the
total set of eighteen responses in Eq.~(\ref{responses}) only twelve survive. 
\item From these twelve responses, the four transferred polarization ones $R^{K'}_{l,s}$ only
contribute when the electron is polarized, while the
four induced polarization ones $R^K_n$ only enter
when FSI are taken into account. 
\end{itemize}
Following the analysis presented in~\cite{Cris1}, in 
this work we limit our attention to those observables that
survive in the plane wave limit, {\em i.e.}, transferred polarization responses 
$R^{TL'}_{l}$, $R^{T'}_{l}$, $R^{TL'}_{s}$, $R^{T'}_{s}$ and transferred asymmetries 
$P'_{l}$, $P'_{s}$. A detailed study of the induced polarization observables 
within RDWIA has been presented in~\cite{Udi00}.

The response functions in Eq.~(\ref{responses}) are constructed directly by taking the appropriate
components of the hadronic tensor $W^{\mu\nu}$ which, within the RDWIA, comes from bilinear combinations 
of the nucleon current matrix elements
\begin{equation}
J^{\mu}_N(\omega,\nq)=\int{d\np\overline{\Psi}_F(\np+\nq)\hat{J}^{\mu}_N\Psi_B(\np)}\, ,
\label{relcurrent}
\end{equation}
where $\Psi_B$ and $\Psi_F$ are 
relativistic wave functions describing the initial bound and final outgoing nucleons, respectively, and
$\hat{J}^{\mu}_N$ is the relativistic one-body current operator.
The bound wave function $\Psi_B$ is a four-spinor with well-defined parity and angular momentum quantum 
numbers $\kappa_b$, $\mu_b$, obtained within the framework of the relativistic independent particle 
shell model. The mean field in the Dirac equation is determined through a Hartree procedure from a 
phenomenological relativistic Lagrangian with scalar (S) and vector (V) terms. It may be written
\begin{equation}
\Psi_B(\np)=\Psi_{\kappa_b}^{\mu_b}(\np)=
        \frac{1}{(2\pi)^{3/2}}\int d \nr e^{-i \np\cdot \nr}
        \Psi_{\kappa_b}^{\mu_b}(\nr)
        =(-i)^{\ell_b}
        \left(\begin{array}{@{\hspace{0pt}}c@{\hspace{0pt}}}
                g_{\kappa_b}(p) \\
                 S_{\kappa_b} f_{\kappa_b}(p)\frac{\nsigma\cdot \np}{p}
                \end{array}\right)\Phi_{\kappa_b}^{\mu_b}(\widehat{\np}) \, 
\label{bwf}
\end{equation}
with $\Phi_{\kappa_b}^{\mu_b}(\widehat{\np})$ the usual spinor harmonics. 
The wave function for the ejected proton $\Psi_F$ is a scattering solution of a Dirac-like equation, 
which includes S-V global optical potentials obtained by fitting elastic proton scattering data. 
This wave function, obtained as a partial wave expansion, is given in momentum space by
\begin{equation}
\Psi_F({\np})=4\pi\sqrt{\frac{E_F+M}{2E_F}}\sum_{\kappa \mu m}e^{-i\delta^\ast_\kappa}
i^\ell\langle\ell m \frac{1}{2} s_F|j \mu\rangle Y_\ell^{m\ast}(\hat{\np}_F)\Psi_\kappa^{\mu}({\np}) \, ,
\end{equation}
where $\Psi_\kappa^{\mu}({\np})$ are four-spinors of the same form as in Eq.~(\ref{bwf}), 
but the phase-shifts and radial functions are complex because of the complex optical potential involved.

Finally, for the nucleon current operator we consider the two choices denoted as CC1 and CC2~\cite{For83}
\begin{eqnarray} 
\hat{J}^\mu_{CC1} &=& (F_1+F_2)\gamma^\mu-\frac{F_2}{2M_N}( 
\overline{P}+P_F)^\mu \label{eqcc1} \\ 
\hat{J}^\mu_{CC2} &=& F_1\gamma^\mu +i\frac{F_2}{2M_N}\sigma^{\mu\nu}Q_\nu  \, ,
\label{eqcc2}
\end{eqnarray} 
where $F_1$ and $F_2$ are the Dirac and Pauli nucleon form factors related to the electric and magnetic 
Sachs form factors in the usual form. The variable $\overline{P}^\mu$ in Eq.~(\ref{eqcc1}) 
is the four-momentum of the 
initial nucleon for on-shell kinematics, {\em i.e.}, $\overline{P}^\mu= 
(\overline{E},\np)$ ($\overline{E}=\sqrt{\np^2+M^2}$ and $\np=\np_F-\nq$).


\subsection{Dynamical effects: projected approach and effective momentum approximation}

In recent years a considerable effort has been devoted to the analysis
of quasielastic $(e,e'p)$ reactions using a fully relativistic formalism. Within this framework, particular
emphasis has been placed on comparison between relativistic and non-relativistic approaches, trying to
identify and disentangle clearly the ingredients which lead to different results in the two types of calculations.
In some recent works~\cite{Meucci01} relativistic effects have been analyzed by comparing directly 
results obtained from a standard non-relativistic DWIA code (DWEEPY) with those provided by a relativistic
calculation. These investigations were aimed at providing systematic and precise information 
on the magnitude of the effects introduced by relativity when compared with the standard 
non-relativistic description
based on DWEEPY. The latter was widely used in the 1980s to analyze low-energy experimental data. However,
although interesting, this study did not allow one to identify clearly the role played by the 
various ingredients entering into
the relativistic formalism. Note that apart from the four-spinor versus two-spinor structure involved
in relativistic and non-relativistic calculations, respectively, also the potentials used in the Dirac
and Schr\"odinger equations for the bound and scattered nucleon are different. Moreover, the non-relativistic
current operator results from an expansion in a basis of free nucleon plane waves and a
Pauli reduction with the operator expanded in powers of $p/M_N$, $q/M_N$ and/or $\omega/M_N$, $p$ being the
missing momentum, $q$ and $\omega$ the transfer momentum and energy, respectively. In
this work we focus on the separate analysis of the various ingredients that enter in the general
formalism, and evaluate their impact on the transferred polarization observables. Hence, in order to
minimize the mismatch coming from the different assumptions involved in 
relativistic and non-relativistic approaches, all of the results presented in this work have been evaluated
using the same potentials and code.
 
Dynamical effects arise from the differences between relativistic and non-relativistic potentials and
wave functions. A detailed study on this subject has been already presented 
in~\cite{Udi99,Udi01,Cab98a}, so here
we simply summarize the basic concepts needed for later discussion of the results. As is well known,
interacting Dirac wave functions have a non-zero overlap with the Dirac sea~\cite{SeWa86}.
The presence of the S-V potentials leads to a significant
dynamical enhancement of the lower components of the Dirac solution at the nuclear interior.
This fact is clearly illustrated by realizing that for a general solution of the Dirac equation with
scalar and vector potentials, its upper and lower components are related by
\begin{equation}
\Psi^{down}=\frac{\nsigma\cdot\np}{E+M_N+S-V}\Psi^{up}
\end{equation}
with $S<0$ and $V>0$. Note that these lower components are enhanced with respect to the ones
corresponding to free positive energy spinors where $S=V=0$. This effect has been referred to as
dynamical enhancement of the lower components, and more recently
as {\it spinor distortion}~\cite{Kel97}.

The analysis of these dynamical effects can be done by constructing properly normalized four-spinor
wave functions where the negative-energy components have been projected out. Thus, instead of the fully
relativistic expression given in Eq.~(\ref{relcurrent}), the nucleon current is evaluated as
\begin{equation}
J^{\mu (+,+)}_N(\omega,\nq)=\int{d\np\overline{\Psi}_F^{(+)}(\np+\nq)\hat{J}^{\mu}_N\Psi_B^{(+)}(\np)}
\, ,
\label{projcurrent}
\end{equation}
where $\Psi_B^{(+)}(\np)$, ($\Psi_F^{(+)}(\np)$) is the positive-energy projection of 
$\Psi_B(\np)$, ($\Psi_F(\np)$), {\em i.e.,}
\begin{eqnarray}
\Psi_B^{(+)}(\np)&=&\Lambda_{(+)}(\np) \Psi_B(\np) \nonumber \\
\Psi_F^{(+)}(\np+\nq)&=&\Lambda_{(+)}(\np+\nq)\Psi_F(\np+\nq)\, ,
\label{projswf}
\end{eqnarray}
where $\Lambda_{(+)}(\np)=(M_N + \overline{\Pbar})/2M_N$ is the positive energy projector.
Then the effects due to the dynamical enhancement of the lower components show up clearly by comparing
the results obtained using the fully relativistic amplitude given in Eq.~(\ref{relcurrent}) with
those evaluated by using Eq.~(\ref{projcurrent}).

Notice that the relationship between lower and upper components 
in the projected wave functions is similar to that corresponding to
free nucleon wave functions, but with the positive-energy projectors depending explicitly on the
integration variable $\np$. An additional approach, referred to as
asymptotic projection, consists of introducing the asymptotic values of the momenta into the positive-energy
projectors acting on the bound and scattered wave functions. This
asymptotic projection is very similar (although it is not completely equivalent) to the effective
momentum approximation (EMA-noSV) introduced originally by Kelly~\cite{Kel97}. Within the EMA-noSV
approach, the four spinors used have 
the same upper components as those of the Dirac equation solutions, but the lower components are 
obtained by enforcing the ``free'' relationship between upper and lower components and using
the asymptotic momenta at the nucleon vertex. Note that these wave functions also lack 
the dynamical enhancement of the lower components.

Finally, one also has the dynamical quenching of the upper component of the Dirac wave function
in the nuclear interior compared with the non-relativistic solution. This effect, associated with the
Darwin term, is implicitly included in
all calculations presented in this work. Hence the differences between the
EMA-noSV approach (or equivalently the asymptotic projection) and the fully relativistic calculation
can be solely ascribed to the negative-energy components.
   
\subsection{Kinematical effects: semi-relativistic reductions}

Another ingredient which leads to differences between the relativistic and 
non-relativistic approaches concerns the specific form of 
the current operator used to evaluate Eq.~(\ref{relcurrent}). Instead of the fully relativistic
operator considered in RDWIA, truncated expressions up to first or higher orders in
$p/M_N$, $\omega/M_N$ and/or $q/M_N$ are employed in standard non-relativistic DWIA calculations.
These effects, here referred to
as kinematical relativistic effects~\cite{Udi99,Udi01,Cris2},
include not only the relativistic kinematics of the nucleon energies 
and momenta~\cite{Ama03,Ama02}
(which must be accounted for in order to describe properly the form of the momentum distribution), but also the effects linked to the use of the relativistic nucleon current operator.

Improved non-relativistic expansions of the nucleon current operator, 
denoted as semi-relativistic approaches, 
which contain important aspects of relativity, have been derived
recently and are available in the literature~\cite{Ama96,Ama96bis,Ama98,Jesch}.  
In this paper we investigate the kinematical effects associated with these
expansions in polarized $(\vec{e},e'\vec{p})$ observables.
To this end we have also incorporated the
semi-relativistic expressions in the relativistic code, so that a direct comparison between
the fully relativistic calculation and the semi-relativistic approach becomes more meaningful because
the effects due to the choice of wave functions and/or potentials are minimized.

To make the analysis clearer, in what follows we explain in some detail the 
procedure used to get the semi-relativistic results.
In the case in which spinor distortion is neglected and asymptotic momenta are used, the  
relativistic ($4\times 4$) current matrix element can be recast in an equivalent form that involves
an effective ($2\times 2$) current operator $\overline{J}^\mu_{eff}$ that occurs between
the upper two component spin $\frac{1}{2}$ spinors.
The ($2\times 2$) operator $\overline{J}^\mu_{eff}$ is obtained
without any approximation concerning non-relativistic reductions; it corresponds to
an exact expression for the on-shell electromagnetic current operator~\cite{Jesch}. This means
that the results obtained using $\overline{J}^\mu_{eff}$ between bispinors corresponding to the
upper components of the relativistic wave functions should coincide exactly with
those obtained using the original relativistic ($4\times 4$) electromagnetic
current operator within the EMA-noSV approach~\cite{Kel97}.
Finally, a comparison between these results and those provided
by making use of the semi-relativistic expressions for the operator, leads to direct information
on the magnitude associated with the kinematical relativistic effects. It is important to point
out that the semi-relativistic reduction is done in the context of the effective momentum approximation,
{\em i.e.,} using asymptotic momenta. 

The semi-relativistic (SR) expression of the electromagnetic current operator relies on the direct
Pauli reduction method, by expanding only in the missing momentum ($p$) over the nucleon mass. 
The transfer energy and momentum are treated exactly. Up to
first-order in $p/M_N$, the following results for the electromagnetic current operators are
obtained:
\begin{equation}
\overline{J}^0=\frac{\kappa}{\sqrt{\tau}}G_E+\frac{i}{\sqrt{1+\tau}}
\left(G_M-\frac{G_E}{2}\right)(\nkappa\times\neta)\cdot\nsigma \, ,
\label{rhosr}
\end{equation}
\begin{eqnarray}
\overline{\nJ} &=& \frac{1}{\sqrt{1+\tau}} \left\{ iG_M(\nsigma\times\nkappa) + 
\left(G_E+\frac{\tau}{2}G_M\right)\neta+G_E\nkappa \right.
 \nonumber \\
&-& \left. \frac{G_M}{2(1+\tau)}(\nkappa\cdot\neta)\nkappa-\frac{iG_E}{2(1+\tau)}
(\nsigma\times\nkappa)\nkappa\cdot\neta \right.
 \nonumber \\
&-& \left. i\tau\left(G_M-\frac{G_E}{2}\right)
(\nsigma\times\neta)+\frac{i(G_M-G_E)}{2(1+\tau)}(\nkappa\times\neta)\nsigma\cdot\nkappa\right\} \, ,
\label{jvecsr}
\end{eqnarray}
where we have introduced the usual dimensionless variables:
$\tau=|Q^2|/4M_N^2$, $\nkappa=\nq/2M_N$ and $\neta=\np/M_N$. Obviously, when computing response functions,
evaluated by taking bilinear combinations of the electromagnetic current matrix elements, terms of order
$\eta^2$ should be dismissed.

As shown, the spin-orbit part of the charge and the relativistic
correction to the transverse current, the first-order convective spin-orbit term, are 
included in Eqs.~(\ref{rhosr},\ref{jvecsr}). Although the above
expressions have been already presented in the 
literature~\cite{Ama96,Ama96bis,Ama98,Jesch,Ama02}, in most of these previous works the analysis of the observables has
been performed adopting additional approximations on the vector current, namely,
$\overline{\nJ}$ is simply taken as the standard non-relativistic reduction except for a global kinematical
factor $(1+\tau)^{-1/2}$
that includes relativistic corrections coming from the Dirac spinors (see~\cite{Ama96,Ama96bis,Ama98,Jesch}
for details). Here we evaluate the recoil nucleon polarized observables by making use of the full
SR currents in Eqs.~(\ref{rhosr},\ref{jvecsr}) taken between the upper components of the original
relativistic wave functions.



\section{Results and discussion}

In this section we analyze the recoil nucleon transferred
polarization observables for proton knockout from $^{16}$O. Although we
focus on results for the $1p_{1/2}$ shell, similar conclusions are reached for 
the $1p_{3/2}$ and $1s_{1/2}$ shells 
unless otherwise specified. Results are computed for both CC1 and CC2 choices of the current 
operator in Eqs.~(\ref{eqcc1},\ref{eqcc2}), and the Coulomb gauge is assumed. A detailed study on 
gauge ambiguities in RPWIA has been presented in Ref.~\cite{Cris1} showing that the Coulomb and Landau gauges
lead to very similar results, differing significantly from the ones corresponding to the Weyl gauge.
These results are proven to persist within the relativistic distorted approach.
The bound nucleon wave function is obtained 
using the parameters of the set NLSH~\cite{Sharma93}. Results computed with other parameterizations
are found to be similar and do not change the general conclusions.
For the outgoing nucleon wave function, we use the energy-dependent,
A-independent potential derived by Clark {\it et al.} for $^{16}$O (EDAIO)~\cite{Coo93} which
describes fairly well the existing elastic proton-$^{16}$O scattering data. Although our main interest
in this work concerns the effects introduced by dynamical and kinematical relativistic effects,
a brief study of the sensitivity of the polarized observables 
to the description of final-state interactions is also presented.
Hence in next section, results evaluated with 
different relativistic optical potentials are shown and compared.
Finally, the Coulomb distortion of the electron wave functions is accounted for by using the
effective momentum approximation with the nuclear Coulomb potential equal to 3.5 MeV
(see~\cite{Udi93,Udi95} for details).
All the results shown throughout this work correspond to the nucleon form factor
parameterization of Gari and Krumplemann~\cite{Gar85}. 

\subsection{Final-State Interactions: relativistic optical potentials}

\begin{figure}[tb]
{\par\centering \resizebox*{0.8\textwidth}{0.6\textheight}{\rotatebox{0}
{\includegraphics{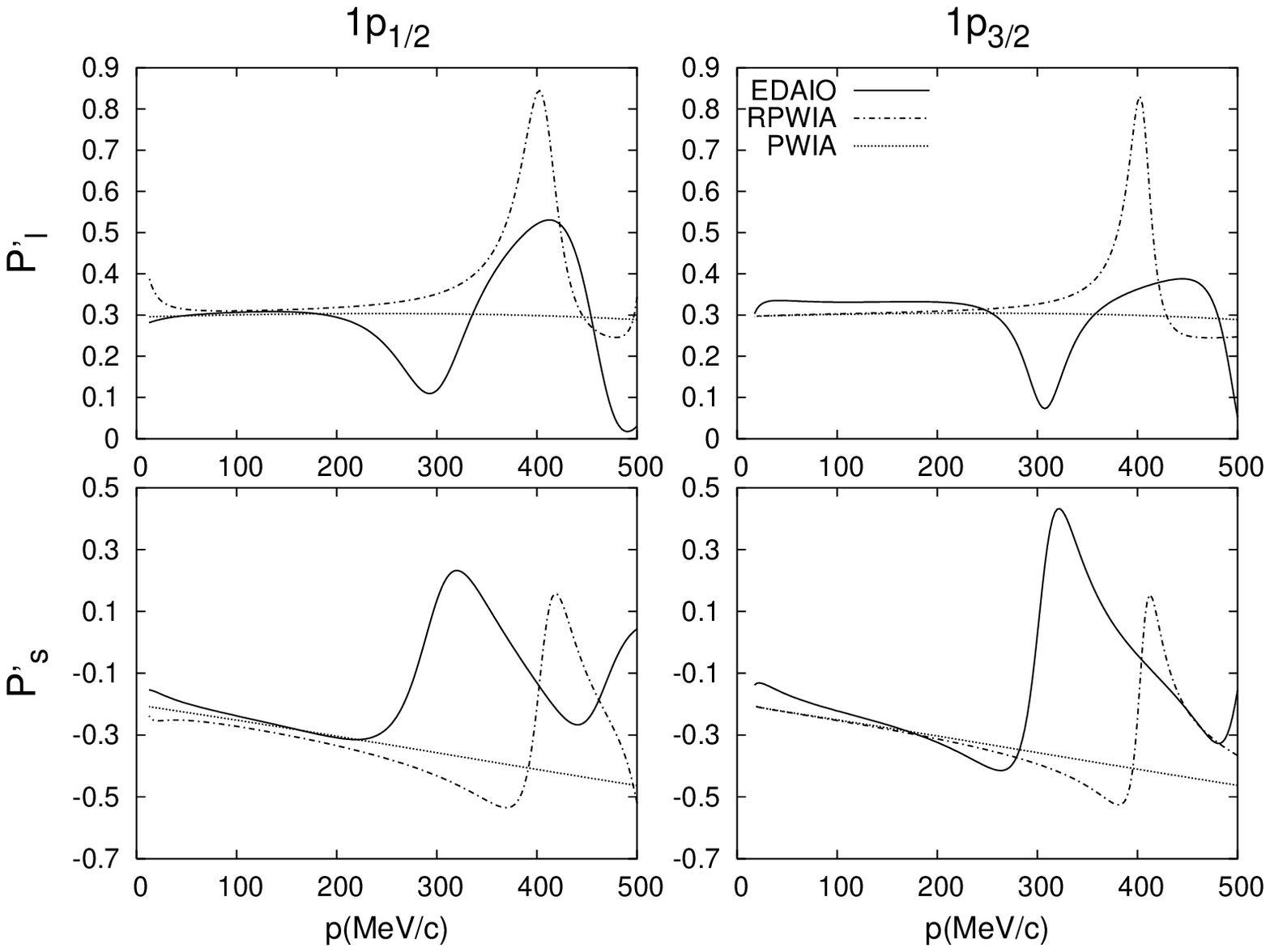}}} \par}
\caption{Transferred polarization asymmetries for the 
\protect\( p_{1/2}\protect \) (left panels) and \protect\( p_{3/2}\protect \) (right panels) shells
in $(q,\omega)$-constant kinematics (see text).
Top and bottom panels correspond to the longitudinal and sideways components, respectively. 
RPWIA results (dot-dashed lines) are compared with RDWIA calculations using EDAIO (solid lines), and with
the PWIA (dotted line) (see text for details). All calculations correspond to the CC2 current
operator.  
\label{fig1}}
\end{figure}

We start our discussion with 
the analysis of the longitudinal and sideways transferred polarization 
asymmetries and their dependence on FSI.
In Fig.~\ref{fig1}, $P'_l$ and $P'_s$ are presented as functions of the
missing momentum $p$. The kinematics are chosen with 
$(q,\omega)$ constant,  $q=1$ GeV/c and $\omega=439$ MeV, yielding $|Q^2|=0.8$ (GeV/c)$^2$.  
This roughly corresponds to the experimental 
conditions of experiments E89-003 and E89-033 performed at JLab~\cite{E89003,E89033,Gao}.
Left panels correspond to the $p_{1/2}$ shell and right panels to $p_{3/2}$.
In each case, RDWIA results obtained with the EDAIO optical potential parameterization~\cite{Coo93} 
are compared with the RPWIA results. Plane wave calculations after projecting out
the negative-energy components of the bound nucleon wave function, denoted as PWIA, are
also shown. Note that PWIA polarization transfer asymmetries coincides
with what one would obtain using free Dirac spinors wave functions for both 
nucleons in Eq.~(\ref{relcurrent}). 
The electron beam energy has been fixed to $\varepsilon_{beam}=2.445$ GeV which
corresponds to an electron scattering angle $\theta_e=23.4^{\circ}$ (forward scattering).

First note the difference between the RPWIA calculations (dot-dashed lines) and the RDWIA results (solid lines).
For low missing momentum values $p\lesssim 200$ MeV/c, the effects of FSI do not modify substantially
the behaviour of the polarization asymmetries, particularly for $P'_l$. However, in the case of $P'_s$,
the difference is of the order of $20$--$25\%$ for $p\simeq 100$ MeV/c which corresponds to the
momentum where the responses reach their maxima for the
$p_{1/2}$ shell. Similar comments also apply to the results obtained
for the $p_{3/2}$ and $s_{1/2}$ shells, although in these cases a smaller effect of FSI is observed
for $P'_s$. It is important to point out that FSI lead to a significant reduction
of the individual response functions:
$\sim$50--60$\%$ ($R^{TL'}_l$) and $\sim$25$\%$ ($R^{TL'}_s$ and $R^{T'}_l$) at
$p\simeq 100$ MeV/c. The response $R^{T'}_s$ is very small and its contribution to
the transferred polarization is hardly visible.
Hence, the results in Fig.~\ref{fig1} clearly
indicate that for low $p$-values, FSI effects are partially cancelled when constructing the transferred
polarization asymmetries. Note also that, for these low-$p$ values, the PWIA approach is more in accord
with the RDWIA. This means that in RPWIA the role of dynamical relativity 
stands out more clearly.

For high missing momentum, $p\gtrsim 200$ MeV/c, FSI strongly modify the behaviour of the polarizations, 
which is in accord
with the peculiar sensitivity to the interaction presented by each response function.
When comparing RDWIA with RPWIA we see that the main
effect is a global displacement to lower momenta of the polarization profiles. Let us 
recall that the oscillatory behaviour shown by $P'_l$ and $P'_s$ within RPWIA is a direct consequence
of the dynamical enhancement of the lower components in the bound Dirac wave functions~\cite{Cris1};
thus disappearing within PWIA. The oscillations
are also present in the relativistic distorted wave calculations, 
although being very different from the RPWIA results with the maxima and minima located at 
different $p$-values. Let us note that the oscillatory behaviour of the 
polarization asymmetries persists even when non-relativistic distorted wave approaches are assumed
(see~\cite{Far03,Meucci01,Ryck99}). This outcome emerges due to the
fact that both FSI and dynamical relativistic effects cause a breakdown of factorization.
A study of the later is presently in progress and the results will be presented in a 
forthcoming publication~\cite{Cab03}.

\begin{figure}[tb]
{\par\centering \resizebox*{0.8\textwidth}{0.6\textheight}{\rotatebox{0}
{\includegraphics{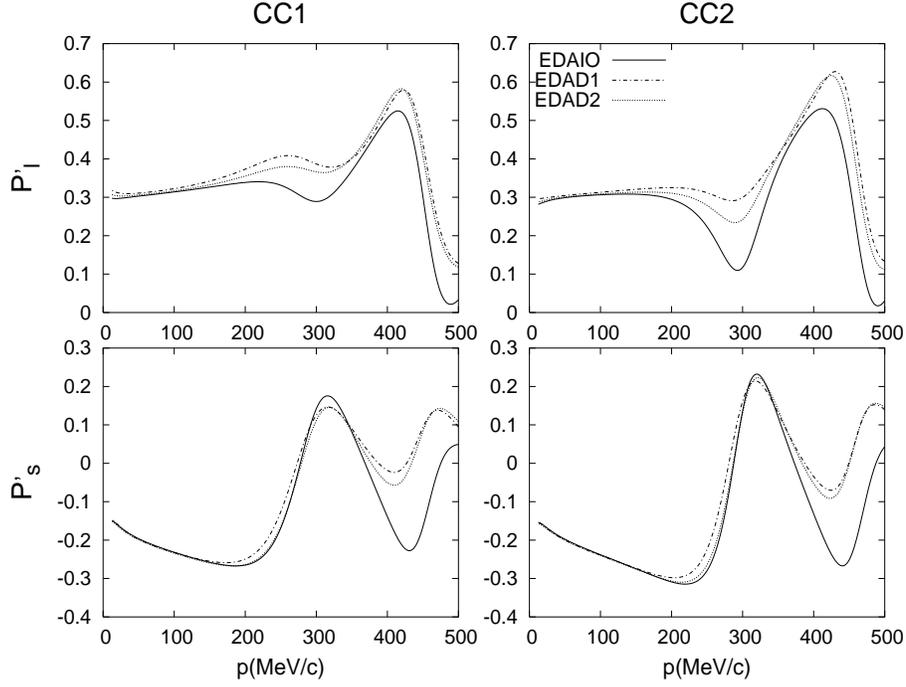}}} \par}
\caption{Transferred polarization asymmetries for the 
\protect\( p_{1/2}\protect \) shell in $(q,\omega)$-constant kinematics. 
Top and bottom panels correspond to the longitudinal and sideways components, respectively. 
Right panels refer to results obtained with the CC2 current operator and left ones to the CC1 current. 
RDWIA calculations using EDAIO (solid lines), 
EDAD1 (dot-dashed lines) and EDAD2 (dotted lines) optical potential parameterizations are compared.
\label{fig2}}
\end{figure}

Let us next focus on the analysis of the uncertainties introduced by different relativistic optical 
potentials. In Fig.~\ref{fig2} we present the transferred ratios $P'_l$ and $P'_s$ for the
$p_{1/2}$ shell evaluated using three different relativistic optical potential 
parameterizations: EDAIO, EDAD1 and EDAD2~\cite{Coo93}. Results with EDAD3 
parameterization are practically identical to those obtained with EDAD1 and therefore have not been plotted. 
The left panels
refer to calculations involving the CC1 current operator and right panels to CC2. 
As pointed out in previous papers~\cite{Far03,Kel96,Meucci01,Kel97}, 
transferred polarization asymmetries 
are expected to be relatively insensitive to the choice of optical potential at low missing momenta. This can be seen in Fig.~\ref{fig2}, at least up to
$p= 150$ MeV/c which is where the cross section reaches its maximum 
value~\cite{Gao}. This trend is also followed in the other two shells, $p_{3/2}$ and $s_{1/2}$.

However, as shown in Fig.~\ref{fig2}, $P'_l$ exhibits a strong dependence on the optical potential parametrization, 
resulting in important differences for larger values of the missing momentum:
$\sim$20$\%$ (CC1) and $\sim$40$\%$ (CC2) for $p\simeq 250$ MeV/c.
Note that in this kinematical region the cross section~\cite{Gao} has already
decreased by almost two orders of magnitude with regards to the maximum, 
making measurements of transferred polarization responses very difficult.
This result contrasts with non-relativistic (NR) and semirelativistic (SR)
approaches where the effects introduced by different non-relativistic 
optical potentials are small~\cite{Far03}. Note also that the current operator 
choice, CC1 versus CC2, gives rise to very significant differences in $P'_l$ within this $p$-region, being of 
the same order as those introduced
by the optical potentials. Only for high $p$-values, $p\gtrsim 350$ MeV/c, is the uncertainty associated with FSI 
larger than that due to the choice of current operator. In the case of the sideways 
polarization $P'_s$, 
in general less dependence on the interaction model as well as on the current is seen, which is
more in accord with non-relativistic analyses. 
Finally, notice that for very high momentum values $p\gtrsim 400$ MeV/c, $P'_l$ and $P'_s$ evaluated
with the EDAIO potential deviate from the results corresponding to the EDAD1 and EDAD2
parameterizations.

To end with this discussion, we conclude that both transferred polarization
asymmetries at moderate $p$ values ($p \simeq 100$ MeV/c) are independent 
of the optical potential choice. Increasing $p$ from here, 
each optical potential starts to follow a different curve especially in the case of $P'_l$.
For very high $p$ ($p\gtrsim 350$ MeV/c), both transferred polarizations present
large sensitivity to the choice of optical potential. However, caution 
should be placed on drawing general conclusions from the results given here in this kinematical region because other ingredients beyond 
the impulse approximation, such as meson exchange currents (MEC), 
$\Delta$-isobar, short-range correlations, {\it etc.,} may also play a crucial role.

\subsection{Dynamical relativistic effects}

\begin{figure}[tb]
{\par\centering \resizebox*{0.8\textwidth}{0.7\textheight}{\rotatebox{0}
{\includegraphics{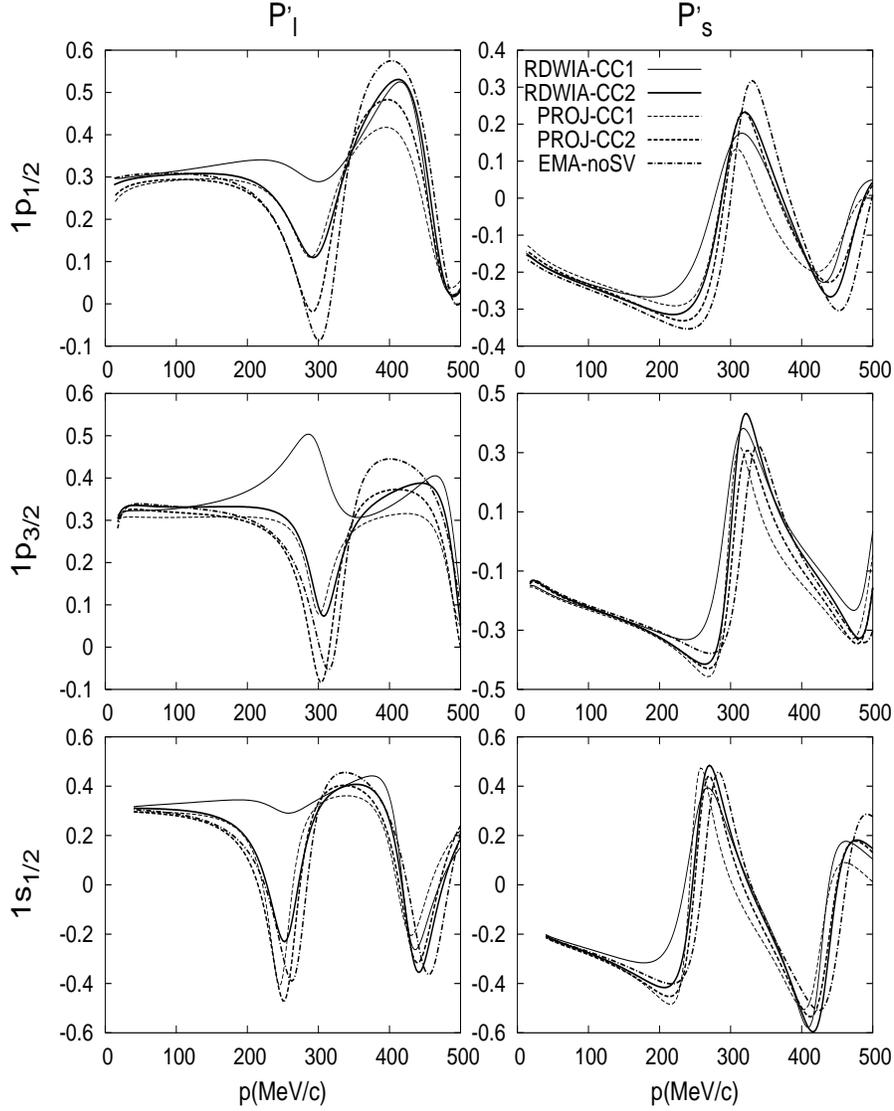}}} \par}
\caption{Same observables as in Fig.~\ref{fig1}. Right panels correspond to $P'_s$ 
and left ones to $P'_l$. On top, middle and bottom panels, results for the $1p_{1/2}$, $1p_{3/2}$ 
and $1s_{1/2}$ shells are plotted, respectively. In each graph, RDWIA calculations evaluated with EDAIO
(solid line) are compared with positive-energy projection results (dashed line) and EMA-noSV approach
(dot-dashed line). Thick lines correspond to the CC2 current operator and thin lines to CC1. 
\label{fig3}}
\end{figure}

This section, which constitutes the main focus of the present work, is devoted to the analysis of dynamical
relativistic effects for nucleon polarized observables within the framework of the RDWIA. With
this aim we present in Fig.~\ref{fig3} the longitudinal and sideways transferred polarization
asymmetries for the three shells involved in $^{16}$O: $p_{1/2}$, $p_{3/2}$ and  $s_{1/2}$.
All of the results have been obtained using the EDAIO optical potential parameterization~\cite{Coo93},
and the choice of kinematics is the same as in the previous figures. To make explicit the effects introduced by
spinor distortion, in each graph we compare the fully relativistic calculations (solid lines) using both
current operators, CC1 (thin lines) and CC2 (thick lines), with the results
after projecting out the negative-energy components 
(see Eqs.~(\ref{projcurrent},\ref{projswf})) (dashed lines). Finally we 
also present for reference the results corresponding to the
EMA-noSV approach evaluated with the CC2 current operator (dot-dashed line). 
Within EMA-noSV, the results provided by the two current operators are very similar,
differing only due to the off-shell kinematical quantities involved in the operator~\cite{Cris2,Cab98a}.

A detailed analysis of the transferred polarizations within the relativistic plane wave approach
was presented in~\cite{Cris1}. In said reference, it is shown that the dynamical
enhancement of the lower components in the bound nucleon wave function leads to strong
oscillations in $P'_{l,s}$ for high missing momentum values, $p\geq 300$ MeV/c. This
behaviour disappears after
projecting out the negative-energy components. From the results shown in Fig.~\ref{fig3}, it is clear that,
within the relativistic distorted wave approximation, the oscillatory behaviour
in the polarization asymmetries persists even after projecting the bound and scattered
proton wave functions over positive-energy states. The same comment
applies to the EMA-noSV approach. On the contrary, this last fact is not applicable to the behaviour shown by the left-right asymmetry $A_{TL}$~\cite{Udi99,Udi01},
defined as the difference of unpolarized cross sections evaluated at $\phi=0^{\circ}$ and $\phi=180^{\circ}$ 
divided by their sum. These results are
connected with the interplay between polarization degrees of freedom and dynamical relativistic
effects. Whereas in RPWIA, projecting out the negative-energy components of the bound nucleon 
wave function leads
to factorization, hence destroying the oscillatory behaviour in $P'_{l,s}$, 
in RDWIA factorization breaks down even after projection over positive-energy components.

From inspection of Fig.~\ref{fig3}, and in accord with previous results for unpolarized 
observables~\cite{Udi99,Udi01,Cab98a} and polarized ones in RPWIA~\cite{Cris1}, we
note that dynamical relativistic effects are maximized for the CC1 current operator. This applies
to both polarization ratios and the three shells considered. Particularly noteworthy is 
the behaviour displayed by $P'_l$ even at intermediate $p$-values in the case of the fully relativistic
CC1 calculation. This result deviates significantly from the others, modifying even the global shape
of the observable. This contrasts with the situation for $P'_s$ where, apart from the specific
discrepancies introduced by relativity, the five calculations follow the same general oscillatory pattern.
Hence it would be interesting to investigate further this intermediate $p$-region where new 
high quality data on $P'_l$ could make it possible to constrain the theoretical
choices for current operator. 

As shown in~\cite{Udi99,Cab98a}, the contribution from the negative-energy components
to the current are of the
same order as the positive-energy ones with the CC2 operator, whereas with the CC1 choice the
negative-energy terms may become much larger. This explains the much
wider spread shown by the CC1 results, particularly the large effects
introduced by the dynamical enhancement of the lower components in $P'_l$.  
As we will show later, this emerges from the polarized responses that enter in the longitudinal 
polarization in contrast with the sideways case. Note also that the CC1
projected calculations get closer to the CC2 ones and to the EMA-noSV approach.
This may indicate that the CC1 current emphasizes the role played by the
lower components in the wave functions, agreeing with the findings for unpolarized responses~\cite{Udi01}.
Precise comparisons with data would yield  definite conclusions on the
reliability of the various approximations.

Finally, it is also interesting to compare the 
effects arising from dynamical relativity with those due to FSI models. As shown in
Figs.~\ref{fig1}--~\ref{fig3}, $P'_l$ presents the strongest sensitivity to both kinds of effects for
intermediate $p$-values, $200\leq p\leq 350$ MeV/c. This can make it difficult to isolate 
the role played by each ingredient when compared with data; however, note that the important deviation 
between the results obtained with the two currents tends to persist, no matter which optical potential is
used. Hence, precise measurements of $P'_l$ in this $p$-region, in conjunction with $P'_s$ data,
may give us important clues to constrain final-state interactions and the 
choice of current operator.



\begin{figure}[tb]
{\par\centering \resizebox*{0.8\textwidth}{0.7\textheight}{\rotatebox{0}
{\includegraphics{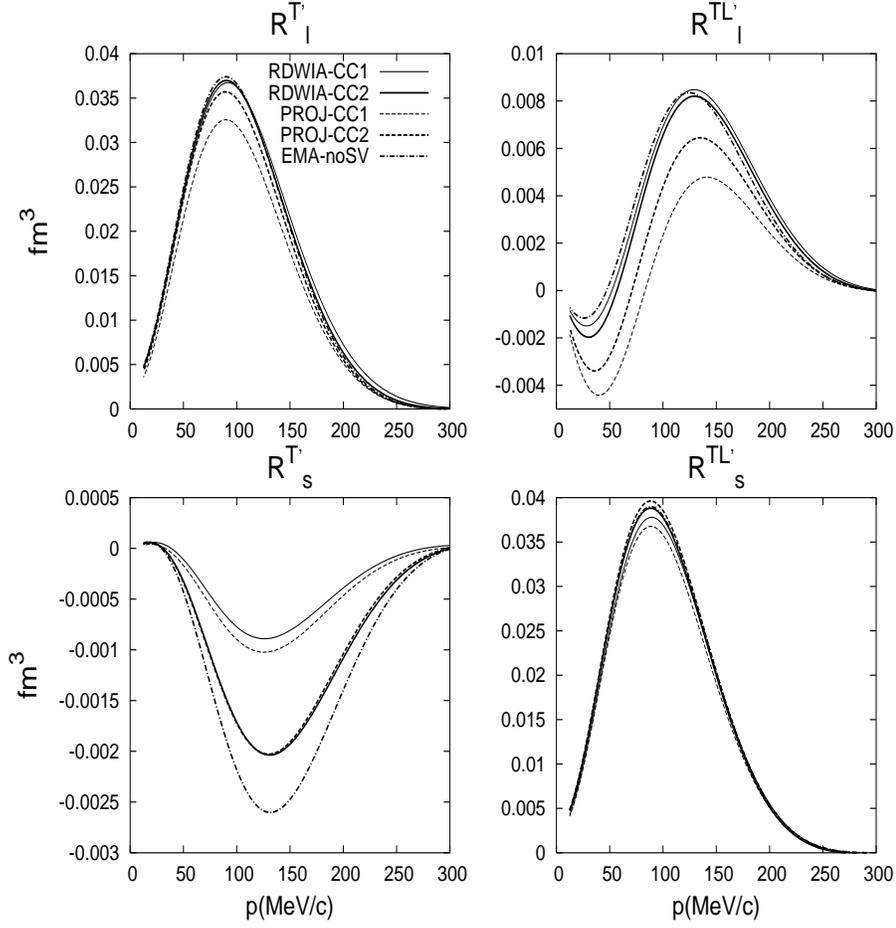}}} \par}
\caption{\label{fig4} Transferred polarized responses for the $1p_{1/2}$ shell. Same kinematics as
in previous figures, and the labelling as in Fig.~\ref{fig2}.}
\end{figure}

To complete the analysis of dynamical relativistic effects, we focus on
the four separate responses that contribute when 
the polarization of the outgoing nucleon is measured and the electron beam is polarized: 
$R^{T'}_l$, $R^{TL'}_l$, $R^{T'}_s$ and $R^{TL'}_s$ ($R^{TL'}_n$ does not enter 
for coplanar kinematics). Results are shown in Fig.~\ref{fig4} for proton knockout in
$^{16}$O from the $p_{1/2}$ shell. Let us recall that Coulomb distortion of the electron waves
breaks the simplicity of Eq.~(\ref{responses}), leading to responses which also depend on the electron
kinematic variables. However, the effective momentum approximation for the electrons adopted in this
work makes Eq.~(\ref{responses}) reliable when analyzing the response functions.
For $^{16}$O we have proven~\cite{Udi01} that Coulomb distortion effects, and consequently the dependence 
of the responses with $\theta_e$, are very small.

As a general rule we observe that $R^{T'}_s$ and $R^{TL'}_l$ show the highest sensitivity to
relativistic dynamics, while the uncertainties in $R^{T'}_l$ and $R^{TL'}_s$ are much smaller. This 
coincides with the analysis already performed in RPWIA~\cite{Cris1} and, although not shown here for
simplicity, applies also to the $p_{3/2}$ and $s_{1/2}$ shells. In addition,
Gordon ambiguities are also significantly enhanced for $R^{T'}_s$ and $R^{TL'}_l$. Finally, note that
the largest spread due to relativistic dynamical effects arises for the CC1 current operator, which is
in accord with RPWIA results~\cite{Cris1}, and can be traced back to the strong influence of the
negative-energy projections of the wave functions in this case.

Let us study in more detail each individual response. As shown in Fig.~\ref{fig4},
the contributions of $R^{T'}_l$ and $R^{TL'}_s$ are rather similar, and moreover,
the EMA-noSV predictions almost coincide (evaluated at the maxima) with the fully 
relativistic calculations, the largest difference being of the order of 3.6\% for the CC1 
current in $R^{TL'}_s$. Positive-energy projected results also follow the RDWIA curves closely, although
sizeable differences are observed for the CC1 current, particularly in the case of $R^{T'}_l$
($\sim$11$\%$ at the maximum).

Concerning $R^{TL'}_l$, we observe that the projected calculations differ substantially 
from the RDWIA results, especially for the CC1 current operator. This 
resembles the large relativistic dynamical effects shown by this response in RPWIA~\cite{Cris1}. 
On the contrary, it is interesting to note that different choices of the current operator within RDWIA
lead to very similar results, which is somewhat opposed to the situation observed in the plane wave
limit~\cite{Cris1}. Finally, the EMA-noSV approach provides a description of $R^{TL'}_l$ 
that basically coincides with the two RDWIA calculations, the largest difference being 
observed at very low-$p$ values. In fact, this result is proven to be valid only 
at $q=1$ GeV/c, where the effective 
momentum approach (EMA) applied to the bound wave function, leads to effects which cancel almost
exactly those coming from the ejected nucleon. 
For lower values of $q$ this cancellation does not occur, and so
an important discrepancy between the EMA-noSV prediction and the RDWIA calculations emerges.

The smallest $R^{T'}_s$ response presents a large dependence on the 
current operator choice. This applies to the full RDWIA calculation as well as to the positive-energy
projected approach. Note, however, that the difference between RDWIA and projected results is 
tiny, almost negligible for the CC2 current. Contrary to $R^{TL'}_l$ case, the EMA-noSV approach for 
$R^{T'}_s$ deviates significantly from the fully relativistic and projected results,
the uncertainty spread (significantly enhanced for the CC1 current) being even larger than that obtained
in RPWIA~\cite{Cris1}. We should also recall that $R^{T'}_s$
is strongly affected by the choice of the optical potential (results corresponding to the
parameterizations EDAIO and EDAD2 are very different from those for EDAD1 and EDAD3).
Although not shown in the figure, it is also important to point out that at low $q$ ($q\leq 350$ MeV/c),
the projection over positive-energies
in the bound nucleon wave function clearly dominates, while
at higher $q$, the reverse occurs. This result contrasts with the behaviour seen for the unpolarized
observables and also with the other three polarized responses, where for high enough transfer momentum
projecting out the negative-energy components in the ejected nucleon wave function is proven not 
to alter the fully relativistic predictions.


The behaviours presented by the four polarized responses, their relative contributions and their
sensitivity to dynamical relativistic effects, give us important clues to understanding the results
obtained for the longitudinal and sideways transferred polarization asymmetries.
The large effects introduced by relativity in $P'_l$, particularly when comparing
full relativistic and projected calculations for CC1, can be traced back to the similar
contributions given by the two responses, $R^{T'}_l$ and $R^{TL'}_l$, that enter in $P'_l$.
Although relativistic dynamics affect $R^{TL'}_l$ more, their effect
on $R^{T'}_l$ is also sizeable. The case of $P'_s$ is clearly different. Here the two polarized
responses involved contribute very differently, $R^{T'}_s$ being much smaller (more than one order
of magnitude). Therefore, the asymmetry $P'_s$ is almost given uniquely by $R^{TL'}_s$, whose
uncertainty due to dynamical relativistic effects presents the lowest spread.
Although results for $p_{3/2}$ and $s_{1/2}$ show basically similar behaviour to those of
the $p_{1/2}$ shell, off-shell and dynamical relativity play a less significant role
for the $p_{3/2}$ shell in $R^{T'}_s$ and $R^{TL'}_l$.

\begin{figure}[tb]
{\par\centering \resizebox*{.9\textwidth}{0.6\textheight}{\rotatebox{270}
{\includegraphics{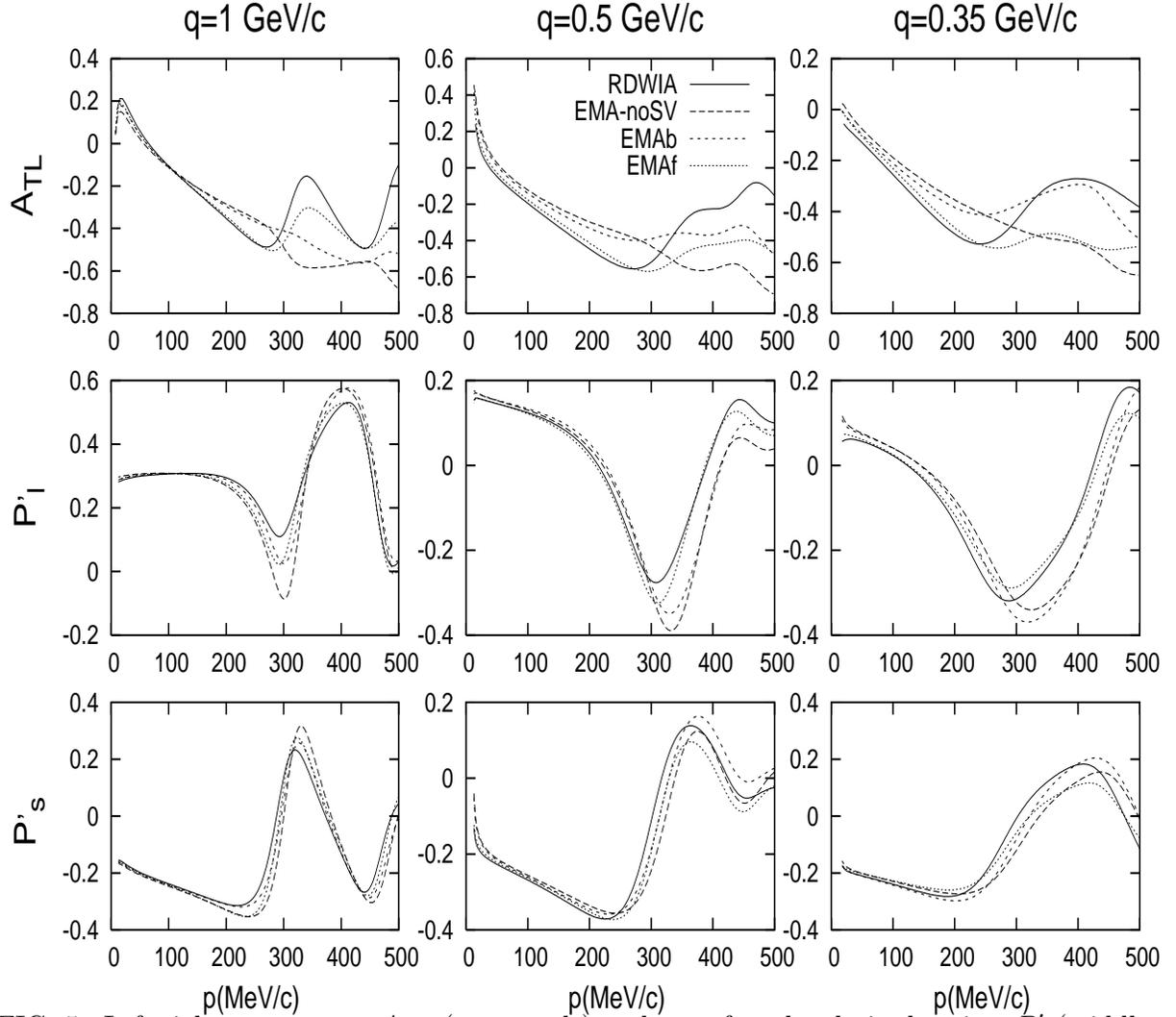}}} \par}
\caption{\label{fig5} Left-right asymmetry $A_{TL}$ (top panels) and transferred polarized ratios,
$P'_l$ (middle panels) and $P'_s$ (bottom panels) for proton knockout from $^{16}$O for the $p_{1/2}$ shell.
Results correspond to $(q,\omega)$-constant kinematics with $q=0.35$ GeV/c (right panels), $q=0.5$ GeV/c
(middle panels) and $q=1$ GeV/c (left panels). In each case the transfer energy $\omega$ is
fixed to the quasielastic peak value. RDWIA calculations (solid line) are compared with the
EMA-noSV approach (dashed line) and with the results after projecting over positive-energy states for the
bound nucleon wave function only (short-dashed line) and for the ejected nucleon only (dotted line).}
\end{figure}

As already mentioned, in RDWIA spinor distortion affects both the bound and ejected nucleon
wave functions. Hence in what follows, we analyze the role of dynamical relativity, isolating the
spinor distortion contribution in each nucleon wave function separately. We show results for
the ratios $P'_{l,s}$ and the left-right asymmetry $A_{TL}$, focusing on the
CC2 current, which minimizes dynamical effects, and the $p_{1/2}$ shell. 
Results for $p_{3/2}$ and $s_{1/2}$ follow the same general trends,
but with a significant reduction of the effects due to relativistic dynamics. 
In Fig.~\ref{fig5} we show the observables for three values of
the momentum transfer $q$. In each case, quasiperpendicular kinematics 
($q$, $\omega$ constant) have been selected, and
RDWIA and projected calculations are compared. Within the projected results, we distinguish 
the EMA-noSV approach, where negative-energy components of the bound and scattered nucleon
wave functions have been projected out, from the results where the projection over positive-energy
components affects only one of the nucleon wave functions: bound (referred to as EMAb) and
ejected (EMAf). 

From inspection of Fig.~\ref{fig5}, a clear difference emerges in the 
behaviour observed for $A_{TL}$ and the polarized ratios $P'_{l,s}$. The asymmetry 
$A_{TL}$ presents a well established pattern: for low-medium $p$-values 
the largest effect shows up when projection over positive-energy states in the bound nucleon 
wave function is assumed (a consequence of the dominance of the direct 
term in the reaction mechanism for low-$p$). On the contrary, for high-$p$ ($p\geq 250,300$ MeV/c), 
the separate influence of each nucleon wave function depends very much on 
$q$. At very low-$q$ the most sizeable effects correspond to
projection of the ejectile wave function state. 
However, as $q$ increases so does the ejected nucleon
momentum $p_F$; thus FSI effects are expected to be smaller and consequently
the contributions of the negative-energy states in the ejected nucleon play a minor role. 
As noted, the results for $P'_{l,s}$ do not match this general behaviour, and it is hard
to state which nucleon wave function plays the major role concerning relativistic dynamical effects.

Finally, a basic difference between $A_{TL}$ and $P'_{l,s}$ connects 
with the oscillatory behaviour shown by these observables. 
While it remains in $P'_{l,s}$ for all $q$-values and all approaches, in the case of $A_{TL}$,
the oscillations disappear when projection is assumed. This effect,
connected with factorization breakdown, is analyzed in~\cite{Cab03}.


\subsection{Semi-relativistic reductions}

\begin{figure}[tb]
{\par\centering \resizebox*{0.8\textwidth}{0.8\textheight}{\rotatebox{0}
{\includegraphics{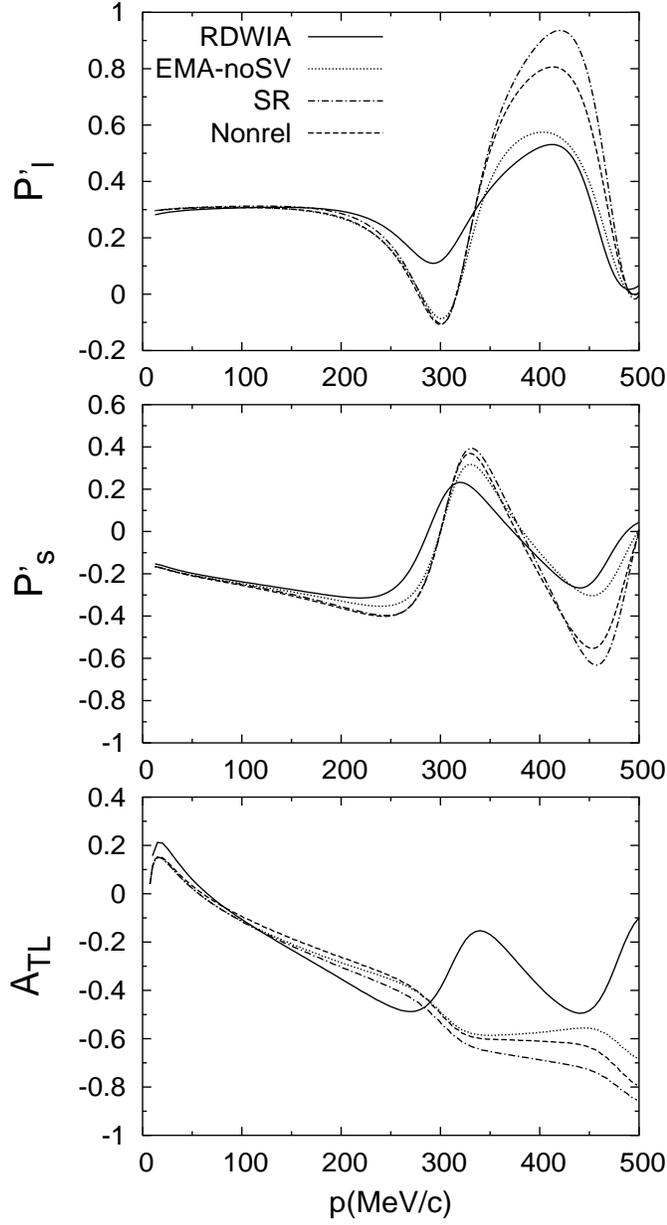}}} \par}
\caption{\label{fig6}Transferred polarizations $P'_l$ (top panel) and $P'_s$ (middle panel), and
$A_{TL}$ asymmetry (bottom panel) for proton knockout from the $p_{1/2}$ shell in $^{16}$O. 
Results correspond to the RDWIA calculation with the CC2 current operator
(solid line), the EMA-noSV approach (dotted line),
the semi-relativistic (SR) current given in Eqs.~(\ref{rhosr},\ref{jvecsr}) 
(dot-dashed line) and the Nonrel approach 
(dashed line) (see text for details). All curves have been obtained using the
EDAIO optical potential.}
\end{figure}

In this section we focus on the kinematical relativistic effects, {\em i.e.,} effects associated
with the non-relativistic reduction of the nucleon current operator. 
In Fig.~\ref{fig6} we present the polarization ratios and $TL$ asymmetry for the $p_{1/2}$ shell
and same kinematics as in Fig.~\ref{fig1}. We compare the RDWIA results (solid line) with the 
EMA-noSV (dotted line) and semi-relativistic approaches. For the latter 
we distinguish the following: SR (dot-dashed line), corresponding to the expressions in
Eqs.~(\ref{rhosr},\ref{jvecsr}),
and Nonrel (dashed line) where additional approximations on the
vector current have been assumed (see Section II.C and \cite{Ama98,Ama02} for details).
As shown, the semi-relativistic curves follow the shape of the EMA-noSV ones, particularly for $A_{TL}$ where oscillations
are largely supressed within EMA-noSV and semi-relativistic approaches. 
Kinematical effects are observed by comparing EMA-noSV and SR calculations. 
As expected, they are very small in the low-p region, increasing
for high missing momenta. This same general pattern emerges for other
transfer momentum values and similar conclusions hold for the
$p_{3/2}$ and $s_{1/2}$ shells.

\begin{figure}[tb]
{\par\centering \resizebox*{0.65\textwidth}{0.7\textheight}{\rotatebox{0}
{\includegraphics{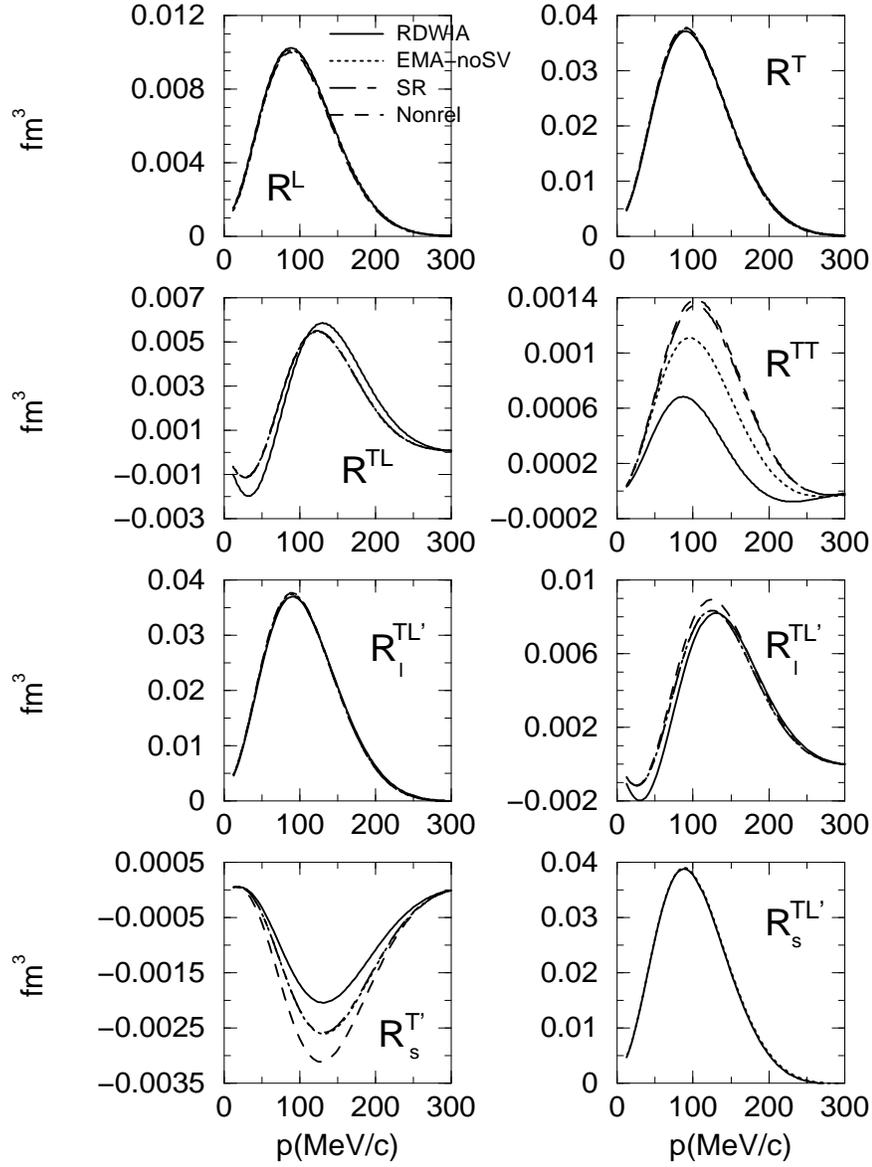}}} \par}
\caption{\label{fig7}Unpolarized (top panels) and recoil nucleon polarized (bottom panels) responses
for proton knockout from the $p_{1/2}$ shell in $^{16}$O. Kinematics as in Fig.~\ref{fig1} 
and the same labelling as in Fig.~\ref{fig6}.}
\end{figure}

To complete the analysis of kinematical effects we study the individual responses. First, let us
consider the unpolarized ones, which are presented in Fig.~\ref{fig7} (top panels) for the $p_{1/2}$ shell
and CC2 current operator. The labelling of the curves is as in previous figure.
We observe that the pure longitudinal and transverse responses, $R^L$ and $R^T$, hardly show any
dependence on either kinematical or dynamical relativistic effects. This coincides with some
previous findings~\cite{Udi99,Udi01}, but clearly disagrees with the results obtained by 
Meucci and collaborators~\cite{Meucci01}, who found very different results for $R^T$ using relativistic and non-relativistic approximations. Concerning $R^{TL}$, it shows a significant
dependence with relativistic nucleon dynamics. This is in accord with our previous analyses~\cite{Udi99,Cab98a},
and also with the results of the Pavia group~\cite{Meucci01}, although in this latter case, the behaviour
found for $R^{TL}$ within the RDWIA calculation, clearly differs from ours for very low
missing momentum. Moreover, notice that the difference between EMA-noSV, SR and Nonrel is
negligible. Finally, the response $R^{TT}$ also shows a high sensitivity to 
both dynamical and kinematical relativistic
ingredients, though its smallness makes it difficult to isolate from cross section
measurements. Let us also recall that our results do not match
those obtained by the Pavia group, particularly for high $q$-values.

Focusing on the transferred polarized responses (bottom panels of Fig.~\ref{fig7}),
we observe that relativistic
ingredients play a very minor role in $R^{TL'}_s$ and  $R^{T'}_l$. On the contrary,
dynamical relativitic effects are sizeable for $R^{TL'}_l$ and especially for
$R^{T'}_s$, while the kinematical relativistic effects are strongly cancelled. Notice that
the EMA-noSV and semi-relativistic approaches give rise to almost identical results. Additional restrictions on the
non-relativistic procedure to get the current operator~\cite{Ama96,Ama96bis,Ama98,Ama02} (Nonrel approach) 
leads to more visible effects which increase when the transferred energy goes up.

\subsection{Comparison with experimental data}

\begin{figure}[tb]
{ \par\centering \resizebox*{0.7\textheight}{0.7\textwidth}{\rotatebox{-90}
{\includegraphics{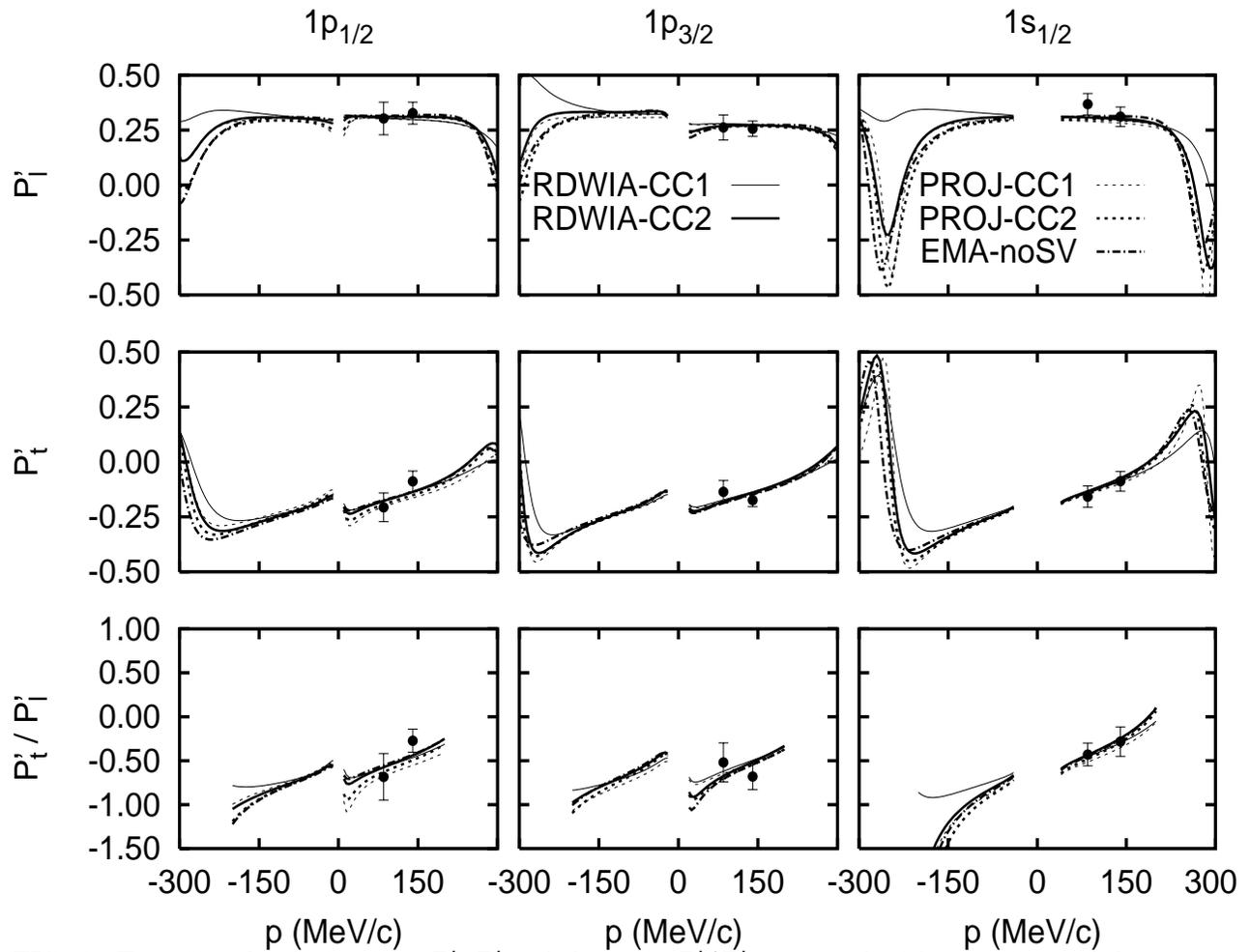}}} \par}
\caption{\label{fig8}Transferred polarizations $P'_l$, $P'_s$ and the ratio $P'_s/P'_l$ compared with
experimental data. Positive (negative) $p$-values refer to $\phi=180^{\circ}$ ($0^{\circ}$). The labelling of the 
curves is as in Fig.~\ref{fig3}. Left, middle and right panels correspond to $p_{1/2}$, $p_{3/2}$ and
$s_{1/2}$ shells, respectively.}
\end{figure}

We proceed to compare our calculations with the experimental data recently measured
at JLab~\cite{Mal00}. The kinematics of the experiment was the same as used in previous
figures except that the azimuthal angle was $\phi=180^{\circ}$ instead of $\phi=0^{\circ}$. As shown later,
this makes an important difference concerning the effects introduced by relativistic dynamics
and/or optical potentials. Fig.~\ref{fig8} shows $P'_l$ (top panels), $P'_t$ (middle panels)
and the ratio $P'_t/P'_l$ (bottom panels) for proton knockout in $^{16}$O from the
$1p_{1/2}$ (left panels), $1p_{3/2}$ (middle panels) and $1s_{1/2}$ (right panels) shells. Note the change of notation for the transverse polarization transfer
observable. In reference ~\cite{Mal00}, and only for $\phi=180^{\circ}$,
the vector perpendicular to the plane containing $\np_F$ and the transfer
momentum $\nq$, is chosen in the opposite direction that we have made in
this paper. Consequently there is also a change of sign in the transverse
vector. In order to present the experimental data taken in~\cite{Mal00} in
the same form as in the original paper, we have prefered to show our
curves for $P'_t$ polarization in Fig.~\ref{fig8}. $P'_t$ is equal to
$P'_s$ for $\phi=0^{\circ}$ and differs only in a sign with $P'_s$ when
$\phi=180^{\circ}$. Curves corresponding to RDWIA, positive-energy projected and EMA-noSV calculations are presented.
The labelling is as in Fig.~\ref{fig3}, and all of the results have been obtained using the EDAIO potential.

To make explicit the differences between $\phi=0^{\circ}$ (kinematics assumed in the previous figures) and
$\phi=180^{\circ}$ (kinematics of the experiment), in each graph we present the polarized observables
as functions of the missing momentum, whose range goes from $-300$ MeV/c to $+300$ MeV/c. Positive
$p$-values refer to $\phi=180^{\circ}$, where the two experimental data are located,
and negative ones to $\phi=0^{\circ}$.

As shown in Fig.~\ref{fig8}, all theoretical calculations satisfactorily reproduce the data, 
improving somehow the general agreement compared with previous SR analyses~\cite{Far03}. However,
it is hard to draw specific conclusions concerning the reliability of the various approaches
within this low-$p$ region.  For higher $p$, relativistic dynamics, off-shell effects and FSI start to
play an important role. In this sense, from inspection of Fig.~\ref{fig8}, it is interesting 
to point out that choosing $\phi=0^{\circ}$
clearly enhances dynamical relativistic effects for $P'_l$ at intermediate $p$-values,
$p\simeq 200-300$ MeV/c. The same comment applies to off-shell and FSI effects. Hence, high quality
$P'_l$ data measured for coplanar, $\phi=0^{\circ}$ kinematics at intermediate $p$-values can provide
precise information to constrain the theoretical models. In the case of $P'_t$, dynamical uncertainties
(also off-shell and FSI effects) are shown to be rather similar for both coplanar
$\phi=0^{\circ}$ and $180^{\circ}$ kinematics.


\subsection{Effects of medium modified form factors}
To finish, we present a brief analysis of the effects introduced by possible changes in the nucleon form factors in the nuclear medium. We limit our attention to the same kinematics as in previous sections. A more exhaustive analysis ranging over different $Q^2$ values, where the models predict different sensitivity to in-medium effects, will be presented in a forthcoming publication. 

The procedure we have used to include these effects in our calculations is as follows. We have taken density-dependent form factors as predicted by the quark-meson coupling model (QMC)~\cite{Lu98}, computed for a bag radius of 0.8 fm. In order to get well behaved modified form factors in the free case, we have scaled the ones parametrized by Gari and Krumplemann~\cite{Gar85} (labelled as $GK$) with the ratio between the QMC form factors at a given density and those predicted for free conditions,
\be
G_{E,M}(Q^2,\rho(\nr))=G_{E,M}^{GK}(Q^2)\frac{G_{E,M}^{QMC}(Q^2,\rho(\nr))}{G_{E,M}^{QMC}(Q^2,0)},
\ee
where $G_{E,M}^{QMC}(Q^2,\rho(\nr))$ are the density-dependent Sachs form factors of the proton immersed in nuclear matter with local baryon density $\rho(\nr)$. By analogy with the free case, we define density-dependent Dirac and Pauli form factors related to $G_{E,M}(Q^2,\rho(\nr))$. Finally, we compute the current matrix elements in coordinate space by introducing these modified form factors into Eqs.~(\ref{eqcc1}) and (\ref{eqcc2}), evaluated for the corresponding local density in $^{16}O$.

\begin{figure}[tb]
{ \par\centering \resizebox*{0.6\textheight}{0.6\textwidth}{\rotatebox{270}
{\includegraphics{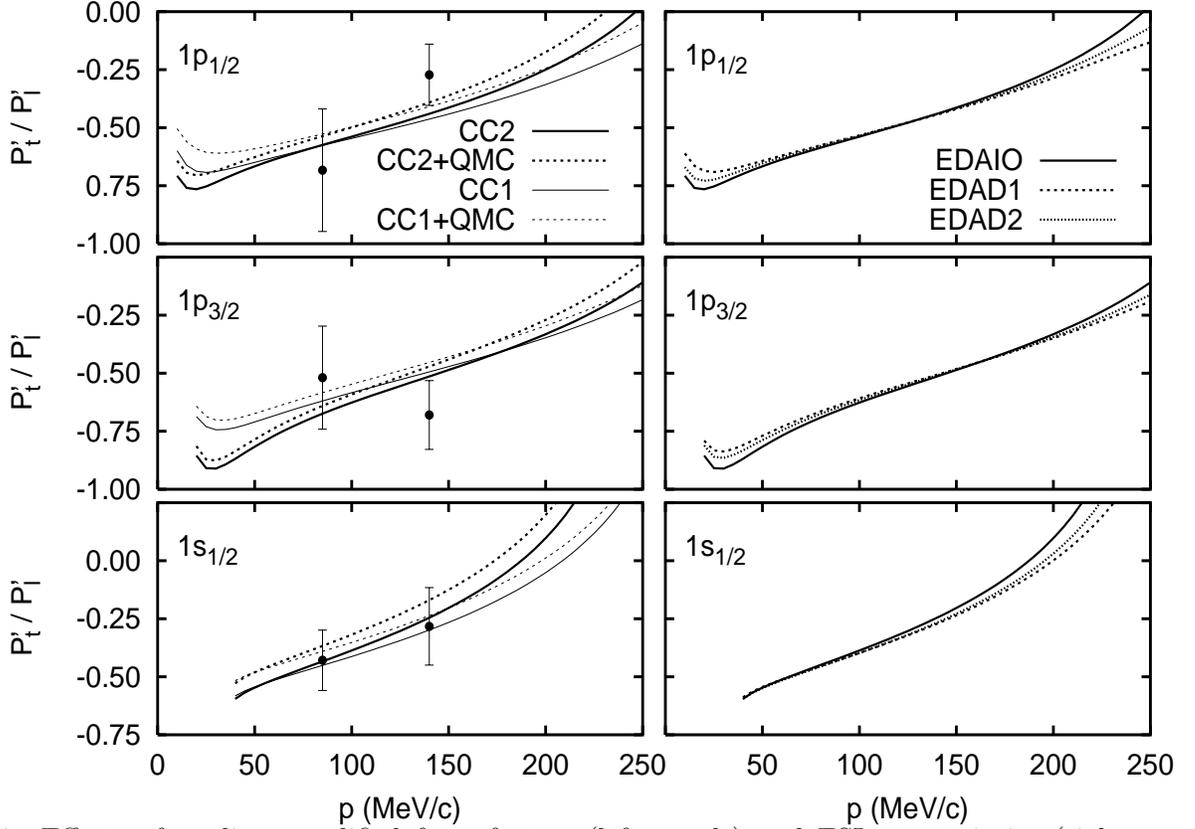}}} \par}
\caption{\label{fig9} Effects of medium modified form factors (left panels) and FSI uncertainties (right panels) on the transferred polarization ratio $P'_t/P'_l$. Results correspond to $\phi=180^{\circ}$. Upper, middle and bottom panels represent the results for $1p_{1/2}$, $1p_{3/2}$ and $1s_{1/2}$, respectively. For the left panels, the free (medium modified) results calculated by using the EDAIO optical potential are represented by solid (dashed) lines. Thick (thin) lines refer to the CC2 (CC1) results, respectively. For the right panels all of the curves have been obtained using CC2. Solid lines correspond to the EDAIO results, dashed lines to EDAD1 and dotted to EDAD2.}
\end{figure}

The results obtained for the ratio of transferred polarization asymmetries are presented in Fig.~\ref{fig9} for both current operators. Only the $\phi=180^{\circ}$ region, where data have been measured, is analysed. As in the previous section, we plot $P'_t/P'_l$ instead of $P'_s/P'_l$. The upper, middle and bottom panels correspond to $1p_{1/2}$, $1p_{3/2}$ and $1s_{1/2}$ knockout, respectively. For completeness, in the right panels of Fig.~\ref{fig9} we also present the uncertainties due to the choice of the optical potential parametrization. As shown, for the $p$ shells our model dependence due to the description of FSI is very small in the region $75 \leq p \leq 175$ MeV/c ($p \leq 100-125$ MeV/c for $1s_{1/2}$), starting to increase for higher $p$. Within this ``safe'' region, medium modification effects for the $p_{1/2}$ amount to $\sim 9\%$ ($\sim 7\%$) for the CC1 (CC2) operator at $p \simeq 100$ MeV/c. Note however that even when these effects are sizeable, the uncertainties introduced by the current operator choice can also be noticeable. The situation worsens for the $1p_{3/2}$ shell, for which the free and QMC calculations get mixed due to the off-shell uncertainties. The precision of the actual experimental data~\cite{Mal00} does not allow one to state which specific calculation is preferred. However, more precise data, particularly in the region $100 \leq p \leq 175$ MeV/c for $p_{1/2}$, could help to constrain the theoretical model. In this sense, note that the QMC results differ more clearly from the free calculations in this shell. 

For the $1s_{1/2}$, the effects of the medium are larger in the vecinity of $p=100$ MeV/c ($\sim 18\%$ for CC2 and $\sim 15\%$ for CC1). Indeed, medium effects are expected to be more important for the inner orbits, due to their higher average densities. The QMC calculations differ substantially from the calculations with free form factors in the $p$ region from 40 to 100 MeV/c, where off-shell ambiguities are very small. In this region it can be possible to 
disentangle density dependence effects if the error bars of the 
data are of the order of $10\%$ or less. At larger $p$ values, off-shell 
ambiguities can make it difficult
 to contrast our predictions including density-dependence of the form factors versus the free ones, as was the case for the $p$-shells. Moreover, other effects beyond the impulse approximation, not considered in this work, could also play an important role in order to provide a precise description of experimental data for the $s$ shell. We have also computed results with other form factor parametrizations (different from the dipole one), and they change the $P'_t/P'_l$ ratio by about $2-3\%$ for both the free and modified case, keeping the relative differences almost unchanged.

In view of these results we conclude that inferring medium modifications from transfer polarization in $^{16}O$ at this $Q^2$ value seems not to be free from ambiguities because of the off-shell effects. However, more precise data and an analysis of other kinematical situations and/or for different nuclei could surely help to draw more definite conclusions.

 
\section{Summary and conclusions}

The analysis of recoil nucleon polarized $(\vec{e},e'\vec{p})$
observables presented in~\cite{Cris1} 
within RPWIA has been extended here to include FSI described through
relativistic optical potentials. The study is restricted to proton knockout from the
$p_{1/2}$, $p_{3/2}$ and $s_{1/2}$ shells in $^{16}$O and quasi-perpendicular kinematics with $q=1$ GeV/c, which
roughly corresponds to the experimental setting. A comparison with data is provided.

The main focus of this paper is to study the role played by the
dynamical enhancement of the lower components in the bound and scattered nucleon wave functions;
along this line, a systematic investigation on the effects linked to FSI and off-shell descriptions
is also done. We show results evaluated with the two usual choices of the
nucleon current operator, CC1 and CC2, and three different relativistic parameterizations of the
optical potential, EDAIO, EDAD1 and EDAD2. Finally, kinematical relativistic effects, associated with
the non-relativistic truncation of the current operator, are also investigated in detail.
Additional ingredients, such as the different relativistic models to describe the bound nucleon wave function and nucleon
form factors, are seen not to modify our conclusions.

From the results shown in previous sections, we may summarize our basic findings as follows:
\begin{itemize}

\item    FSI constitutes a basic ingredient in order to get reliable results to be compared
  with data. Transferred polarization ratios as well as polarized responses do modify very
  significantly their structure when FSI are taken into account. However, a kind of cancellation
  of the FSI effects is observed to occur in $P'_l$ and $P'_s$ for low missing momenta, 
  $p\leq 100$ MeV/c. Concerning the role of the optical potential, a clear difference emerges
  for the two asymmetries at very high $p$-values, $p\geq 400$ MeV/c, when comparing results for
  the EDAIO and EDAD-type potentials. This is due to the different reduction of the scattered
  wave function in the nuclear interior produced by the two kinds of optical potentials. Finally,
  at intermediate $p$-values ($p\simeq 250$ MeV/c), $P'_l$ shows a strong dependence on the
  interaction model, whereas the uncertainty in $P'_s$ is tinier. A similar comment applies also
  to the off-shell ambiguities.

\item   Dynamical relativistic effects are shown to be very important, being enhanced for the
  CC1 current operator. Concerning the responses, 
  $R^{T'}_s$ and $R^{TL'}_l$ present the highest dependence with dynamical effects, as also found in the
  RPWIA studies. However, contrary to the plane wave limit, where the dynamical enhancement of
  the lower components of the bound nucleon completely modifies the shape of the transferred asymmetries, in the case of the
  distorted wave approach the general oscillatory behaviour of $P'_l$ and $P'_s$ persists even
  after projecting-out the negative-energy components. This differs also with the behaviour of the unpolarized observable $A_{TL}$. This effect is linked to the breakdown of factorization. At intermediate $p$-values,
  $P'_l$ shows a stronger sensitivity to relativistic
  dynamics.

\item   Results corresponding to SR reductions are proven to be very similar (depending on the truncation)
  to the EMA-noSV approach, differing more from the RDWIA calculations. As expected, the difference between the three approaches increases as p goes up. The SR approaches also lead to a significant cancellation
  of the oscillatory behaviour in $A_{TL}$, while maintaning the general shape of $P'_l$ and $P'_s$. This
  is again connected with the factorization property and its possible breakdown.

\end{itemize}
From the comparison with experimental data, we show the
reliability of our general description of $(\vec{e},e'\vec{p})$
reactions, and conclude that new high quality data measured at 
intermediate $p$-values ($150-200$ MeV/c) may help to constrain 
the various theoretical approximations involved in our calculations. 

As pointed out in~\cite{Far03}, other ingredients that go
beyond the impulse approximation, such as those arising from meson exchange
currents and the $\Delta$-isobar contribution,
may also play a very important role in properly describing the
transferred polarization asymmetries. These remain to be investigated in
a relativistic context, although, in on-going work, the inclusion of two-body currents
within the fully relativistic formalism is presently in progress.
In the final analysis, any interpretation in terms of medium modified 
nucleon form factors requires having excellent control of all of these  
model dependences, both those discussed in the present work and those
that go beyond the impulse approximation. Within our model we have found that for the kinematical conditions of E89003 and E89033~\cite{E89003,E89033} it is difficult to separate effects introduced by density-dependent form
factors from off-shell ambiguities due to the choice of current operator. However, for the s1/2 and p1/2 shells there is a region in
between 40 and 100 MeV/c that is relatively free from off-shell
uncertainties and where the effect of medium modifications would
be easier to assess. In a future
publication we will present the results of a more extensive study in the context of
the nuclear model uncertainties and will assess the impact of including
medium modifications of the form factors at different values of $Q^2$. 


\subsection*{Acknowledgements}

This work was partially supported by funds provided by DGI (Spain) and
FEDER funds, under Contracts Nos BFM2002-03315, BFM2002-03562, FPA2002-04181-C04-04 and BFM2000-0600
and by the Junta de Andaluc\'{\i}a (Spain) and
in part by the U.S. Department of Energy under
Cooperative Research Agreement No. DE-FC02-94ER40818.
M.C.M. and J.R.V. acknowledge financial support from the Fundaci\'on
C\'amara (University of Sevilla) and the Consejer\'{\i}a de Educaci\'on de la
Comunidad de Madrid, respectively.





\newpage

\newpage

\newpage

\newpage

\newpage

\newpage

\newpage

\end{document}